\documentclass[preprint,amsmath,amssymb]{revtex4}

\usepackage{graphicx}
\usepackage{dcolumn}
\usepackage{bm}

\begin{document}

\title{Thermal soliton correlation functions in theories with a $Z(N)$ symmetry}

\author{Leonardo Mondaini}
\affiliation{Grupo de F\'{i}sica Te\'{o}rica e Experimental, Departamento de Ci\^encias Naturais, Universidade Federal do Estado do Rio de Janeiro, Av. Pasteur 458, Urca, Rio de Janeiro-RJ 22290-240, Brazil \\ Email: mondaini@unirio.br}


\begin{abstract}
We show that the quantum solitons occurring in theories describing a complex
scalar field in $(1+1)$-dimensions with a $Z(N)$ symmetry may
be identified with sine-Gordon quantum solitons in the phase of this field. Then using both the Euclidean thermal Green function 
of the two-dimensional free massless scalar field in coordinate space and its dual, we obtain an explicit series expression for the corresponding solitonic correlation function at finite temperature.
\end{abstract}


\maketitle


\section{Introduction}

As remarked in \cite{MM2}, the sine-Gordon (SG) model is certainly one of the best studied of (1+ 1)-dimensional physics. 
The interest in this field theoretical model has been
enhanced by its connections with the two-dimensional (2D) neutral Coulomb
gas (CG)\cite{Samuel} and also with the 2D XY-magnetic system \cite{Kosterlitz}. In this
framework, it becomes an useful and powerful tool for the study of a
great variety of physical properties of these two systems, which in principle
admit actual realizations in nature. The SG model is also integrable in the
sense that the spectrum and the S-matrix are exactly known \cite{Zamolod}.

In this work we employ the same methodology established in \cite{MondainiMarinoMPLA} in order to obtain
an explicit series expression for the two-point thermal soliton correlation function in theories with a $Z(N)$ symmetry. This has been done, firstly,
by observing that when we use a polar representation for the complex scalar field  in (1+1)-dimensions described 
by a theory with $Z(N)$ symmetry, we are naturally led to a SG theory in the phase of this field \cite{MondainiMarinoJPA}.
Then using the connection between the SG theory and the 2D neutral CG along with the representation for the relevant soliton creation operators introduced in \cite{ms, mbs, sq2}, and both the Euclidean thermal Green function 
of the 2D free massless scalar field in coordinate space and its dual \cite{MondainiMarinoMPLA, LM}, we obtain our expression for the corresponding solitonic correlation function at finite temperature.

\section{Quantum phase solitons in theories with a $Z(N)$ symmetry}

We start by considering the following theory describing a complex
scalar field in (1+1)-dimensions,
\begin{equation}
\mathcal{L}=\partial_\mu\phi^\ast\partial^\mu\phi+\gamma\left({\phi^\ast}^N
+\phi^N\right)-\eta\left(\phi^\ast\phi\right)^M, \label{5.1.1}
\end{equation}
where $N$ and $M$ are integers and $\gamma$ and $\eta$ are real
parameters. The above Lagrangian density is invariant under the $Z(N)$ transformation:
$\phi(x,t)\rightarrow e^{i\frac{2\pi}{N}}\phi(x,t)$. The choice
$\gamma > 0$, implies the spontaneous breakdown of the $Z(N)$
symmetry. In this case, as is well-known, the theory will have degenerate vacua and
soliton excitations. A full quantum theory of these solitons was
developed in \cite{ms,mbs,sq2}. This includes an explicit expression
for the soliton creation operator, namely
\begin{equation}
\mu(x)=\exp\left\{-\frac{2\pi}{N}\int_{x,C}^{\infty}d\xi_\nu
\epsilon^{\mu\nu}
\phi^\ast(\xi){\stackrel{\leftrightarrow}{\partial}}_\mu
\phi(\xi)\right\}, \label{1.2.2.8}
\end{equation}
and a general expression for its local Euclidean correlation
function,
\begin{widetext}
\begin{equation}
\langle\mu(x)\mu^\dagger(y)\rangle = \mathcal{N}\int
\mathcal{D}\phi^\ast \mathcal{D}\phi \exp\left\{-\int
d^2z\left[\left(D_\mu\phi\right)^\ast
\left(D_\mu\phi\right)+V\left(\phi^\ast,\phi\right)\right]\right\},\label{1.2.2.3}
\end{equation}
\end{widetext}
where
\begin{equation}
D_\mu=\partial_\mu-i\alpha A_\mu, \ \ \ \ \ \ \ \ \ \ \ \
A_\mu(z,C)=\int_{x,C}^{y}d\xi_\nu
\epsilon^{\mu\nu}\delta^2(z-\xi).\label{1.2.2.4}
\end{equation}
In the above expression, $V$ is the potential of an arbitrary
Lagrangian and the integral is taken along an arbitrary curve $C$,
connecting $x$ and $y$. It can be shown, however, that Equation
(\ref{1.2.2.3}) is independent of the chosen curve.

We are going to show in what follows that these quantum solitons may
be identified with SG quantum solitons in the phase of the field
$\phi$.

Using the polar representation for $\phi$, namely,
$\phi(x,t)=\rho(x,t) e^{i\theta(x,t)}$, where $\rho$ and $\theta$
are real fields, we can rewrite the previous Lagrangian as
\begin{equation}
\mathcal{L}=\partial_\mu\rho\partial^\mu\rho+\rho^2\partial_\mu\theta\partial^\mu\theta+2\gamma\rho^N\cos
N\theta-\eta\rho^{2M} . \label{5.1.3}
\end{equation}

As we shall argue, the topological properties
of the theory are not affected by $\rho$
fluctuations. Thus, from now on, we will make the constant $\rho$
approximation:
\begin{equation}
\rho(x,t)=\rho_0, \ \ \ \ \ \ \rho_0 \,\ \textrm{constant}.
\label{5.1.4}
\end{equation}
Substituting the above equation into Equation (\ref{5.1.3}) we obtain
\begin{equation}
\mathcal{L}=\rho_0^2\partial_\mu\theta\partial^\mu\theta+2\gamma\rho_0^N\cos
N\theta-\eta\rho_0^{2M}, \label{5.1.5}
\end{equation}
which, clearly, is a SG Lagrangian in $\theta$.

Indeed, one may observe that the particular case of the potential appearing in the Lagrangian presented in Equation (\ref{5.1.1}) for $N=M=2$
\begin{equation}
V(\phi^\ast,\phi)=-\gamma\left({\phi^\ast}^2
+\phi^2\right)+\eta\left(\phi^\ast\phi\right)^2, \label{5.3.1}
\end{equation}
presents a doubly degenerate vacuum ($Z(2)$ symmetry), indicating the occurrence of spontaneous symmetry breaking.

After rewriting Equation (\ref{5.3.1}), in terms of polar fields, as
\begin{equation}
V(\rho,\theta)=-2\gamma\rho^2\cos 2\theta+\eta\rho^4, \label{5.3.2}
\end{equation}
we can see that there are two degenerate minima situated at the points
\begin{equation}
(\rho,\theta)=(\rho_0,\theta_0)=
\begin{cases}
(\sqrt{\frac{\gamma}{\eta}},0) \\
(\sqrt{\frac{\gamma}{\eta}},\pi)
\end{cases}
\label{5.3.3}
\end{equation}
where the potential assumes the value
$V(\rho_0,\theta_0)=-\gamma^2/\eta$. Applying, then, the constant
$\rho$ approximation
\begin{equation}
\rho(x,t)\ = \rho_0=\sqrt{\frac{\gamma}{\eta}} \label{5.3.4}
\end{equation}
and adding $\gamma^2/\eta$ (such that the potential vanishes at its minima) we get the following SG potential for the
phase field $\theta$
\begin{equation}
V(\theta)=\frac{2\gamma^2}{\eta}\left(1-\cos 2\theta\right).
\label{5.3.5}
\end{equation}
The above potential gives rise, when solving the corresponding Euler-Lagrange equation for the static case
\begin{equation}
\frac{\partial^2 \theta}{\partial x^2}=\frac{\eta}{2\gamma}\frac{\partial V(\theta)}{\partial \theta},
\label{elstatic}
\end{equation}
to the following solutions (classical solitonic excitations)
\begin{equation}
\theta(x)=\pm 2\arctan\{\exp\left[\sqrt{\gamma}\, (x-x_0)\right]\},
\label{5.3.7}
\end{equation}
where the plus and minus signs correspond, respectively, to a soliton
($\theta_s(x-x_0)$) and an anti-soliton \emph{in the phase of the complex scalar field}
$\phi$.

Finally, since
\begin{equation}
\lim_{x\rightarrow -\infty} \theta_s(x-x_0)=0\ \ \ \ \ \
\textrm{and}\ \ \ \ \ \ \lim_{x\rightarrow +\infty}
\theta_s(x-x_0)=\pi, \label{5.3.8}
\end{equation}
we may see that this \emph{phase soliton} connects the minima of the potential presented in Equation
(\ref{5.3.2}) when $\rho=\rho_0 = \sqrt{\gamma/\eta}$, as it should.

From the above considerations, we may then conclude that, in the constant-$\rho$ approximation, the theories
given by Equation (\ref{5.1.1}) will present SG solitons in the phase of the
complex scalar field $\phi$. The corresponding topological current
will be
\begin{equation}
J^\mu=\epsilon^{\mu\nu}\partial_\nu \theta, \label{5.1.5a}
\end{equation}
which gives rise to the following representation for the topological charge operator
\begin{equation}
\begin{split}
\mathcal{Q}&=\int_{-\infty}^\infty dx' \, J^0 = \int_{-\infty}^\infty
dx' \,
\partial_{x'} \theta(x',t) \\ &= \theta(+\infty,t)- \theta(-\infty,t).
\end{split}
\label{1.2.3.2}
\end{equation}

We can see, from the above expression, that topological properties are indeed related to large $\theta$
fluctuations and, therefore, the constant $\rho$ approximation
should not interfere in such properties. Moreover, substituting Equation (\ref{5.3.7}) into Equation (\ref{1.2.3.2}) and making use of Equation (\ref{5.3.8}), we can also see that the value of the topological charge associated to the presented classical solitonic excitations is equal to $\pi$.

From a quantum mechanical point of view, however, in order to describe the quantum states associated with the classical solitonic excitations shown in Equation (\ref{5.3.7}), we need to introduce a quantum soliton creation operator $\mu$, whose application on the vacuum state of our theory, namely $|0\rangle$, yields a solitonic state $|\emph{Soliton}\rangle$, i.e., $\mu|0\rangle=|Soliton\rangle$. This quantum soliton creation operator must satisfy the following commutation relation with the topological charge operator $\mathcal{Q}$, presented in Equation (\ref{1.2.3.2}):
\begin{equation}
\left[\mathcal{Q},\mu\right] = \sigma\mu,
\label{commutation}
\end{equation}
where $\sigma$ is a real constant whose meaning may be clearly understood by applying both sides of the above equation on the vacuum state $|0\rangle$. Indeed, since $\mu|0\rangle=|Soliton\rangle$ and $\mathcal{Q}|0\rangle=0$, we have
\begin{equation}
\left[\mathcal{Q},\mu\right] |0\rangle= \sigma\mu|0\rangle \Rightarrow \mathcal{Q}|Soliton\rangle=\sigma|Soliton\rangle,
\label{eigenstates}
\end{equation}
from which we may conclude that $\sigma$ is nothing but the eigenvalue of the topological charge operator $\mathcal{Q}$ associated to the eigenstate (solitonic state) $|Soliton\rangle$ or, in other words, the expectation value of $\mathcal{Q}$ when measured in the state $|Soliton\rangle$. This property allows us to associate $\sigma$ with the classical topological charge.

Thus, as we have already saw, we may explicitly confirm the fact that the soliton operators
introduced in \cite{sq2,ms,mbs} are indeed creation operators of
quantum solitons in the phase of the complex scalar field $\phi$ by computing the commutation relation between the quantum soliton
creation operator and the topological charge operator.

In order to do that, let us firstly observe that, since the momentum canonically conjugated to $\theta$
is
$$\pi_\theta=\frac{\partial\mathcal{L}}{\partial
\dot{\theta}}=2\rho^2\dot{\theta},$$ we can rewrite the soliton creation operator for the theory described by Equation (\ref{5.1.1}) in terms of polar fields (in Minkowski space) as
\cite{sq2,ms,mbs} 
\begin{equation}
\mu(x,t)=\exp\left\{-i\ \frac{2\pi}{N}\int_{x}^{\infty}d\xi_1
\pi_{\theta}(\xi_1,t)\right\}. \label{1.2.3.3}
\end{equation}

Notice that this is nothing but the Mandelstam creation operator of
quantum solitons in the SG model \cite{mand}, as it should. Then,
using canonical commutation relations along with the well-known Baker-Campbell-Hausdorff identity $\left[A,e^B\right]=\left[A,B\right]e^B$,  we readily find
\begin{equation}
\begin{split}
\left[\mathcal{Q},\mu\right] &=
\left\{-i\frac{2\pi}{N}\int_{-\infty}^\infty dx' \partial_{x'}
\int_{x}^{\infty}d\xi_1 \left[\theta(x',t),
\pi_\theta(\xi_1,t)\right]\right\}\mu
 \\ &= \frac{2\pi}{N}\mu.
\end{split}
\label{1.2.3.4}
\end{equation}

Last but not least, following the above discussion, we may notice that Equation (\ref{1.2.3.4}) implies that the operator $\mu$ creates
eigenstates of the topological charge operator $\mathcal{Q}$ with eigenvalue
$2\pi/N$ (which, not by coincidence, for $N=2$, corresponds to the value of the topological charge associated to the classical solitonic excitations presented in Equation (\ref{5.3.7}), namely, $\sigma=\pi$), thus proving that the quantum solitons occurring in the theory described by Equation (\ref{5.1.1}) are, indeed, SG solitons in the phase of the complex scalar field $\phi$, i.e., they are {\it phase solitons}. 

In the next session, we are going to 
calculate the two-point correlation function of these quantum soliton excitations at finite temperature.

\section{Two-point thermal soliton correlation function}

The Euclidean vacuum functional of the SG theory, for an arbitrary
$T$ may be written as the grand-partition function of a classical 2D CG of point charges $\pm \eta$, contained in an infinite strip of
width $\beta=1/k_BT$, interacting through the potential $G_T({\bf{r}})$, namely \cite{nicola}
\begin{equation}
\begin{split}
\mathcal{Z}& =
\sum_{m=0}^{\infty}\frac{(\gamma\rho_0^N)^m}{m!}\sum_{\{\lambda_i\}_m}\int_0^\beta\int_{-\infty}^{\infty}
\prod_{i=1}^{m} d\tau_i\, dz_i \\
& \quad
\times\exp\left\{-\frac{N^2}{4\rho_0^2}\sum_{i=1}^{m}\lambda_i\sum_{j=1}^{m}\lambda_j
G_T({\bf{z}}_i-{\bf{z}}_j)\right\},
\end{split}
\label{3.6.1}
\end{equation}
where $\lambda_i=\pm 1$, $\sum_{\{\lambda_i\}_m}$ runs over all
possibilities in the set $\{\lambda_1,\ldots,\lambda_m\}$, and
$G_T({\bf{r}})$ is the Euclidean thermal Green function of the 2D free
massless scalar theory in coordinate space (${\bf{r}}\equiv
(x,\tau)$), which is given by \cite{Das}
\begin{equation}
G_T({\bf{r}})=\frac{1}{\beta}\sum_{n=-\infty}^{\infty}\int_{-\infty}^{\infty}
\frac{dk}{2\pi}\frac{e^{-i(kx+\omega_n\tau)}}{k^2+\omega_n^2},
\label{gft1}
\end{equation}
with $\omega_n=2\pi n/\beta$. 

A closed-form representation for this function was presented for the first time in \cite{delepine}, and is given by
\begin{equation}
G_T({\bf{r}})=-\frac{1}{4\pi}\ln\left\{\frac{\mu_0^2\,\beta^2}{\pi^2}\left[\cosh\left(\frac{2\pi}
{\beta}x\right)-\cos\left(\frac{2\pi}{\beta}\tau\right)\right]\right\}.
\label{greenfunction}
\end{equation}
This has also been obtained by using methods of integration on the
complex plane \cite{LM}. At $T=0$, $G_T({\bf{r}})$ reduces to the 2D
Coulomb potential and we retrieve the usual mapping onto the Coulomb
gas \cite{Samuel}.

We may rewrite the above thermal Green function in terms of the new
complex variable
\begin{equation}
\zeta({\bf{r}})\equiv
\zeta(z)=\frac{\beta}{\pi}\sinh\left(\frac{\pi}{\beta}\,z\right),
\label{zeta}
\end{equation}
where  $z=x+i\tau$, as
\begin{equation}
G_T({\bf{r}})=\lim_{\mu_0\rightarrow
0}-\frac{1}{4\pi}\ln\left[\mu_0^2\zeta({\bf{r}})\zeta^*({\bf{r}})\right]. \label{gft20}
\end{equation}

Notice that in the zero temperature limit
($T\rightarrow 0$, $\beta\rightarrow \infty$), we have $\zeta
(z)\rightarrow z $ and $\zeta^*(z)\rightarrow z^* $ and, therefore,
we recover the well-known Green function at zero temperature, namely
\begin{equation}
\lim_{\beta\rightarrow\infty}G_T({\bf{r}};\mu_0)=-\frac{1}{4\pi} \ln
\left[\mu_0^2\, zz^*\right]=-\frac{1}{4\pi} \ln \left[\mu_0^2 ||{\bf{r}}||^2 \right]. \label{gft22}
\end{equation}

We can now determine the soliton correlation function at $T\neq 0$, by
using Equation (\ref{1.2.3.3}) along with the gas representation of
the vacuum functional, Equation (\ref{3.6.1}). Notice that, the insertion of $\mu$ operators corresponds, in the CG language, to the introduction of
``magnetic" fluxes on
the gas \cite{ms}. The soliton correlator, therefore, is nothing but the
exponential of the interaction energy of the associated classical
system. Charges and ``magnetic" fluxes interact with their similar,
through the thermal Green function $G_T({\bf{r}})$, whereas the
charge-flux interaction occurs via the dual thermal Green function
$\tilde{G}_T({\bf{r}})$ (described in the Appendix) \cite{ms}. This is the reason why it is crucial
to know this function in order to obtain the soliton correlator.

Following the above considerations and the same procedure employed
at $T=0$ \cite{MM2}, we can write  the two-point thermal soliton correlation function occurring in theories with a $Z(N)$ symmetry as
\begin{widetext}
\begin{equation}
\begin{split}
\langle&\mu({\bf{x}})\mu^\dagger({\bf{y}})\rangle_T
\\ &=
\mathcal{Z}^{-1}\sum_{m=0}^{\infty}\frac{(\gamma\rho_0^N)^m}{m!}\sum_{\{\lambda_i\}_m}\int_0^\beta\int_{-\infty}^{\infty}
\prod_{i=1}^{m} d\tau_i\, dz_i\,\exp \left\{\frac{1}{2}\int d^2z\,
d^2z' \right . \\ & \quad \left . \times
\left[\left(i N \sum_{i=1}^m \lambda_i\,\delta^2({\bf{z}}-{\bf{z}}_i)+\frac{4\pi\rho_0^2}{N}\left[\int_{{\bf{x}}}^{{\bf{y}}}d\eta_\mu\,
\epsilon^{\alpha\mu}\delta^2({\bf{z}} - {\boldsymbol\eta})\right]\partial_\alpha^{(z)}\right)\right .\right .\\ & \quad
\left .\left .\times\left(i N \sum_{j=1}^m \lambda_j\,\delta^2({\bf{z}}'-{\bf{z}}_j)+\frac{4\pi\rho_0^2}{N}\left[\int_{{\bf{x}}}^{{\bf{y}}}d\xi_\nu\,
\epsilon^{\beta\nu}\delta^2({\bf{z}}' - {\boldsymbol\xi})\right]\partial_\beta^{(z')}\right)\right .\right .\\&\quad \left .\left
.\times \,\left(\frac{1}{2\rho_0^2}\,G_T({\bf{z}}-{\bf{z}}')\right)\right]\right\}.
\end{split}
\label{e3}
\end{equation}
\end{widetext}

After some algebra, we get
\begin{widetext}
\begin{equation}
\begin{split}
\langle&\mu({\bf{x}})\mu^\dagger({\bf{y}})\rangle_T
\\ &=
\mathcal{Z}^{-1}\sum_{m=0}^{\infty}\frac{(\gamma\rho_0^N)^m}{m!}\sum_{\{\lambda_i\}_m}\int_0^\beta\int_{-\infty}^{\infty}
\prod_{i=1}^{m} d\tau_i\, dz_i \,\exp
\left\{-\frac{N^2}{4\rho_0^2}\sum_{i=1}^{m}\lambda_i\sum_{j=1}^{m}\lambda_j
G_T({\bf{z}}_i-{\bf{z}}_j)\right .
\\ & \quad \left .+2 \pi i \sum_{i=1}^{m}\lambda_i\int_{{\bf{x}}}^{{\bf{y}}}d\xi_\nu
\epsilon^{\beta\nu}\partial_\beta^{(\xi)}G_T({\bf{z}}_i-{\boldsymbol\xi})+\frac{4\pi^2\rho_0^2}{N^2}\int_{{\bf{x}}}^{{\bf{y}}}d\eta_\mu\int_{{\bf{x}}}^{{\bf{y}}}d\xi_\nu\,
\epsilon^{\alpha\mu}\partial_\alpha^{(\eta)}\epsilon^{\beta\nu}
\partial_\beta^{(\xi)}\right . \\ & \quad \left . \times \,G_T({\boldsymbol\eta}-{\boldsymbol\xi})\right\}.
\end{split}
\label{e5}
\end{equation}
\end{widetext}

Using the Cauchy-Riemann
conditions, Equation (\ref{cauchyriemann}), we may rewrite the above expression as
\begin{widetext}
\begin{equation}
\begin{split}
\langle&\mu({\bf{x}})\mu^\dagger({\bf{y}})\rangle_T
\\ &=
\mathcal{Z}^{-1}\sum_{m=0}^{\infty}\frac{(\gamma\rho_0^N)^m}{m!}\sum_{\{\lambda_i\}_m}\int_0^\beta\int_{-\infty}^{\infty}
\prod_{i=1}^{m} d\tau_i\, dz_i \,\exp
\left\{-\frac{N^2}{4\rho_0^2}\sum_{i=1}^{m}\lambda_i\sum_{j=1}^{m}\lambda_j
G_T({\bf{z}}_i-{\bf{z}}_j)\right .
\\ & \quad \left .-2\pi i  \sum_{i=1}^{m}\lambda_i\left[\tilde{G}_T({\bf{z}}_i-{\bf{y}})-\tilde{G}_T({\bf{z}}_i-{\bf{x}})\right]-\frac{8\pi^2\rho_0^2}{N^2}\left[G_T({\bf{0}})-G_T({\bf{x}}-{\bf{y}})\right]\right\}.
\end{split}
\label{e5}
\end{equation}
\end{widetext}

Finally, using the UV-regulated version of $G_T({\bf{r}};\mu_0)$, namely
\begin{equation}
G_T({\bf{r}};\mu_0,\varepsilon) = -\frac{1}{4\pi} \ln \left\{\mu^2_0
\left[\zeta({\bf{r}})\zeta^*({\bf{r}})
+|\varepsilon|^2\right]\right\},\label{3.7a}
\end{equation}
and the expression (\ref{gft20c}) for the dual thermal Green
function $\tilde{G}_T({\bf{r}})$, we obtain
\begin{widetext}
\begin{equation}
\begin{split}
\langle&\mu({\bf{x}})\mu^\dagger({\bf{y}})\rangle_T 
\\&=
\lim_{\varepsilon\rightarrow 0}\lim_{\mu_0\rightarrow 0}
{\mathcal{Z}}^{-1}\left[ \frac{|\varepsilon|^2}{\zeta({\bf{x}}-{\bf{y}})\zeta^*({\bf{x}}-{\bf{y}})}\right]^{\frac{2\pi\rho_0^2}{N^2}}\sum_{n=0}^{\infty}\frac{\alpha^{2n}}{(n!)^2}\int_0^\beta\int_{-\infty}^{\infty}
\prod_{i=1}^{2n} d\tau_i\, dz_i
\\ & \quad \times  \exp \left
\{\frac{N^2}{16\pi\rho_0^2}\sum_{i\neq j=1}^{2n}\lambda_i\lambda_j\ln
\left\{\mu^2_0 \left[\zeta({\bf{z}}_i-{\bf{z}}_j)\zeta^*({\bf{z}}_i-{\bf{z}}_j) +|\varepsilon|^2\right]\right\}\right . \\ & \quad \left .+\frac{1}{2}\sum_{i=1}^{2n}
\lambda_i\ln \left[\frac{\zeta({\bf{z}}_i-{\bf{y}})\zeta^*({\bf{z}}_i-{\bf{x}})}{\zeta^*({\bf{z}}_i-{\bf{y}})\zeta({\bf{z}}_i-{\bf{x}})}\right] \right \},
\end{split}
\label{3.8}
\end{equation}
\end{widetext}
where the renormalized coupling $\alpha$ (Coleman's renormalization) is given by $\alpha = (\gamma\rho_0^N)
\left(\mu_0^2|\varepsilon|^2\right)^{N^2/(16\pi\rho_0^2)}$ \cite{col}.

Notice also that, as in the $T=0$ case, existence of the $\mu_0\rightarrow 0$ limit
imposes the neutrality of the gas, namely $\sum_{i=1}^m
\lambda_i=0$, because in this case the $\mu_0$-factors are
completely canceled. This implies that the index $m$ appearing in Equation
(\ref{3.6.1}) must be even ($m=2n$, with $n$
positive and $n$ negative $\lambda_i$'s) and, therefore,
$\sum_{\{\lambda_i\}_m}=(2n)!/(n!)^2$.

Let us finally remark that in the $T\rightarrow 0$ limit the above
correlation function reduce to the corresponding function of the zero
temperature theory \cite{MM2}, as it should.

\section*{ACKNOWLEDGMENTS}


This work has been supported in part by Funda\c c\~ao CECIERJ.


\section*{APPENDIX}

\renewcommand{\theequation}{A\arabic{equation}}

\setcounter{equation}{0}

Here we present a closed-form representation for the dual thermal Green function $\tilde{G}_T({\bf{r}})$, which we have used for computing the soliton correlation function shown in Equation (\ref{e5}).

Indeed, we can see from Equation (\ref{gft20}) that the thermal Green
function may be written as the real part of an analytic function of
the complex variable $\zeta$, namely
\begin{equation}
G_T({\bf{r}};\mu_0)=\textrm{Re}\left[\mathcal{F}(\zeta)\right]=\frac{1}{2}\left[\mathcal{F}(\zeta)+\mathcal{F}^*(\zeta)\right],
\label{gft20b}
\end{equation}
where $\mathcal{F}(\zeta)\equiv -(1/2\pi)\ln\left[\mu_0\zeta({\bf{r}})\right]$.

The imaginary part of $\mathcal{F}(\zeta)$ may be written as
\begin{equation}
\begin{split}
\tilde{G}_T({\bf{r}})\equiv\textrm{Im}\left[\mathcal{F}(\zeta)\right]&=\frac{1}{2i}
\left[\mathcal{F}(\zeta)-\mathcal{F}^*(\zeta)\right]\\&=-\frac{1}{4\pi
i}\ln\left[\frac{\zeta({\bf{r}})}{\zeta^*({\bf{r}})}\right].
\label{gft20c}
\end{split}
\end{equation}

Now, from the analyticity of $\mathcal{F}(\zeta)$, it follows
that its imaginary and real parts must satisfy the Cauchy-Riemann
conditions, which are given by
\begin{equation}
\epsilon^{\mu\nu}\partial_\nu G_T =-\partial_\mu \tilde{G}_T, \ \ \
\ \ \ \ \ \ \ \ \ \epsilon^{\mu\nu}\partial_\nu
\tilde{G}_T=\partial_\mu G_T. \label{cauchyriemann}
\end{equation}
This property characterizes $\tilde{G}_T$ as the dual thermal Green
function.

\end{document}